%                                               macros version 11/05/96
\input epsf

\magnification= \magstep1
\tolerance=1600 
\parskip=5pt 
\baselineskip= 6 true mm \mathsurround=1pt
\font\smallrm=cmr8
\font\smallit=cmti8

\def\secbreak{\vskip12pt plus 1in \penalty-200\vskip 0pt plus -1in} 
 %\prefbreak{distance}
 
\def\a{\alpha}          \def\b{\beta}     
\def\d{\delta}          \def\D{\Delta}  \def\e{\varepsilon}
\def\h{\eta}                        
\def\m{\mu}                             \def\vv{\varphi}
\def\n{\nu}                 
\def\r{\varrho}         \def\s{\sigma}  
            \def\th{\theta}      
                     
\def\w{\omega}

\def\cl{\centerline}    
\def\ni{\noindent}      \def\pa{\partial}       \def\dd{{\rm d}}        
\def\tl{\tilde}                 \def\bra{\langle}       \def\ket{\rangle}

\def\fn#1#2{\footnote{$^#1$} {\scrunch #2 \toe}}
\def\fnd#1{\footnote{$^\dagger$} {\scrunch #1 \toe}}

        \def\scrunch{\baselineskip=10 pt \smallrm}
        \def\toe{\hfil\break\vskip-18pt}
        \newcount\noteno
\def\numfn#1{\global\advance\noteno by 1 
        \footnote{$^{\the\noteno}$} {\scrunch #1 \toe}}
                % numbered footnote

\def\fractt#1#2{{\textstyle{#1\over#2}}}
\def\fract#1#2{\raise .35 em\hbox{$\scriptstyle#1$}\kern-.25em/\kern-.2em\lower .22 em
\hbox{$\scriptstyle#2$}}

\def\half{\fract12} 

\def\part#1#2{{\partial#1\over\partial#2}} 
 \def\ref#1{${\vphantom{)}}^#1$}

\def\bbf#1{\setbox0=\hbox{$#1$} \kern-.025em\copy0\kern-\wd0
        \kern.05em\copy0\kern-\wd0 \kern-.025em\raise.0433em\box0}              
                % boldface in math mode.
\def\qu{\ {\buildrel {\displaystyle ?} \over =}\ }
\def\df{\ {\buildrel{\rm def}\over{=}}\ }

\def\low#1{{\vphantom{]}}_{#1}} 
  
\def\ref#1{${\,}^{\hbox{\smallrm #1}}$}
 
\def\lap{\setbox0=\hbox{$<$}\,\raise .25em\copy0\kern-\wd0\lower.25em\hbox{$\sim$}\,}
\def\glt{\setbox0=\hbox{$>$}\,\raise .25em\copy0\kern-\wd0\lower.25em\hbox{$<$}\,}
\def\gap{\setbox0=\hbox{$>$}\,\raise .25em\copy0\kern-\wd0\lower.25em\hbox{$\sim$}\,}
\def\br{\hfil\break}

\def\Gbar{\raise.13em\hbox{--}\kern-.4em G}
\def\Lbar{\raise.13em\hbox{--}\kern-.5em L} 
\def\Mbar{\setbox0=\hbox{--}\,\raise .27em\copy0\kern-.6em\hbox{$M$}}
\def\Ebar{\setbox0=\hbox{--}\,\raise .27em\copy0\kern-.5em\hbox{$E$}}

%{\nopagenumbers %
\vglue 1truecm
\rightline{THU-96/31}
\rightline{gr-qc/9608037}
\vfil
\cl{\bf QUANTIZATION OF SPACE AND TIME}\smallskip
\cl{\bf IN 3 AND IN 4 SPACE-TIME DIMENSIONS\fn*{Lectures held at the
NATO Advanced Study Institute on ``Quantum Fields and Quantum Space
Time", Carg\`ese, July 22 -- August 3, 1996.}}

\vfil

\cl{G. 't Hooft }
\bigskip
\cl{Institute for Theoretical Physics}
\cl{University of Utrecht, P.O.Box 80 006}
\cl{3508 TA Utrecht, the Netherlands}
\smallskip\cl{e-mail: g.thooft@fys.ruu.nl}
\vfil
\ni{\bf Abstract}
 
{\narrower The fact that in Minkowski space, space and time are both
quantized does not have to be introduced as a new postulate in physics,
but can actually be derived by combining certain features of General
Relativity and Quantum Mechanics. This is demonstrated first in a model
where particles behave as point defects in 2 space dimensions and 1
time, and then in the real world having 3+1 dimensions. The mechanisms
in these two cases are quite different, but the outcomes are similar:
space and time form a (non-cummutative) lattice.

These notes are short since most of the material discussed in these
lectures is based on two earlier papers by the same author
(gr-qc/9601014  and  gr-qc/9607022), but the exposition given in the
end is new.
\smallskip}

\vfil\eject % }\pageno=1 %

{\bf\ni 1. IN 2+1 DIMENSIONS}\medskip

If we remove one space-dimension, Einstein's theory of gravity becomes
a beautiful and simple theory. In the absence of a cosmological
constant, space-time is locally flat, and the simplest matter sources,
point particles, form conical singularities in 2-space. When at rest,
they cause no curvature in the time direction\fnd{If the particle has
{\smallit spin\/} however, the monodromies on curves surrounding them
show a constant jump in time.} Space-time surrounding moving point
particles is understood by performing Lorentz transformations. Quite
generally, space-time can be described by sewing together flat
3-simplexes\ref1.

The rich structure of this apparently very simple model emerges when
one attempts to construct sequences of Cauchy surfaces. It is
convenient to choose these Cauchy surfaces also to consist of simplexes
(polygons) sewn together. At the seams, the surface thus obtained may
be curved, but of course the Riemann curvature of 3-space is still
required to vanish at these seams; it is only non-vanishing at the
location of the point particles.

Within each simplex of the Cauchy surface there is a preferred Lorentz
frame (with the time axis orthogonal to the surface). By choosing time
to run equally fast on all simplexes we define a simple series of
Cauchy surfaces. The polygons glued together evolve according to well
defined rules. Polygons may even split in two, or disappear, and in
each of these cases the further evolution of the Cauchy surface is
uniquely defined\ref2. It can be simulated on a computer\ref3.

\midinsert\epsffile{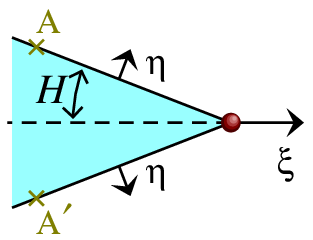}\narrower{\ni\bf Figure~1.} Wedge cut
out by a moving particle (dot). $\xi$ is the boost parameter for the
velocity of the particle; $\eta$ is that for the velocity of the wedge.
The Hamiltonian $H$ is one-half the wedge angle.\endinsert
 
The rules for the evolution of a Cauchy surface have been derived in
Refs\ref{2, 3}. Where there is a particle there is a cusp (Fig.~1),
where the points $A$ and $A'$ must be identified.  When the particle is
at rest we identify (one-half of) the opening angle of the cusp with
the mass $\m$ of the particle. If the particle moves, the cusp must be
oriented in such a way that the direction on the velocity coincides
with the bisectrix of the cusp angle, so as to avoid any time jump
across the cut. The Lorentz contraction formula gives the new angle
$H$, and plain geometry relates the velocity $\tanh\eta$ of the cusp's
edges to the velocity $\tanh\xi$ of the particle:
$$\eqalignno{\tan H&\,=\,\cosh\xi\,\tan\m\,,&(1.1)\cr
\tanh\eta&\,=\,\sin H\,\tanh\xi\,.&(1.2)\cr}$$
Algebraically, one derives from this:
$$\eqalignno{\cos\m&\,=\,\cos H\,\cosh\eta\,,&(1.3)\cr
\sinh\eta&\,=\,\sin\m\,\sinh\xi\,.&(1.4)\cr}$$
These equations are to be compared with the more familiar properties
of particles in flat space-time:
$$\eqalignno{H&\,=\,\m\,\cosh\xi\,,&(1.1a)\cr
p&\,=\,H\,\tanh\xi\,,&(1.2a)\cr
\m^2&\,=\,H^2-p^2\,,&(1.3a)\cr
p&\,=\,\m\,\sinh\xi\,.&(1.4a)\cr}$$

\midinsert\epsffile{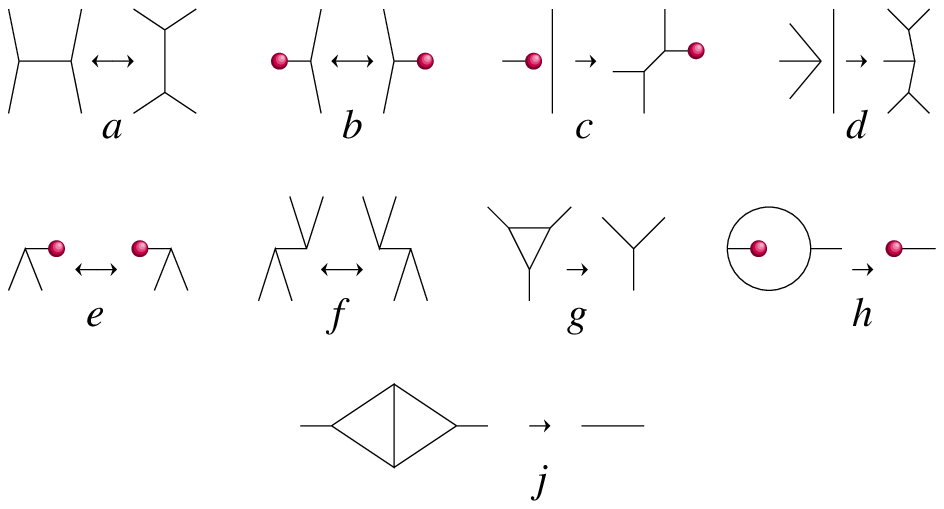}\narrower\ni{\bf Figure 2.} The nine distinct
transitions that can occur among the polygons, indicated diagrammatically.
\endinsert

At a vertex between three polygons $I$, $II$, and $III$, one must note
that the Lorentz boost from $I$ to $III$ can be written as the product
of the boost from $II$ to $III$ and the one from $I$ to $II$. This
gives us relations between the velocities of the edges of the adjacent
polygons and their angles\ref2. Because the Cauchy surface is not flat,
the three angles at one vertex need not add up to $2\pi$. We write
$$\a_1+\a_2+\a_3\,=\,2\pi-2\w\,.\eqno(1.5)$$ The nine different
possible polygon transitions are indicated diagrammatically in Fig.~2.
It turned out to be instructive to study the classical cosmological
models obtained with a limited number of particles. The space-time
topology is typically chosen to be $S_2\times R_1$, but one can take
also higher genus surfaces for the spacelike component. Depending on
the initial state chosen, the final state of the ``universe" is found
to be in one of two possible classes:\br $i$) an indefinitely expanding
universe, in which the edges of all polygons continue forever to
increase in length. Eventually, everything goes radially outwards, and
no further transitions take place. Or:\br $ii$) the universe continues
to shrink, faster and faster. There is a natural end point at a time
$t_{\rm end}$ at which it shrinks to a point. Before that time is
reached, however, an infinite number of transitions have taken place,
and each particle sees all the other particles pass at ever decreasing
impact parameters (transverse separation distances). The speed at which
they pass each other, in the center of mass frame, approaches
exponentially that of light. A typical final state is depicted in
Fig.~3.

\midinsert\epsffile{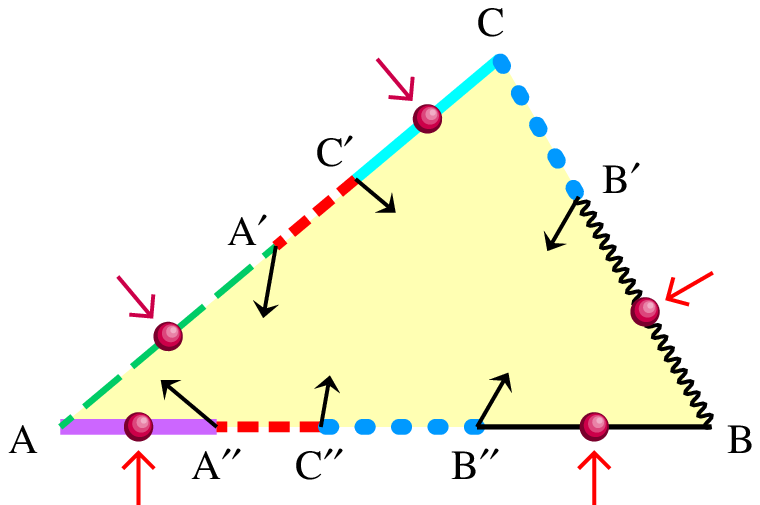}\narrower\ni{\bf Figure~3.} Example of a
shrinking final state of a universe. The particles have so large $\xi$
values that all wedges opened up to form angles of practically
$180^\circ$. They all move inwards, nearly with the speed of light
(arrows outside frame). Edges of equal texture in the picture are to be
matched. $A$, $A'$ and $A''$ are to be identified; similarly $B$, $B'$,
$B''$, and $C$, $C'$ and $C''$, respectively. The vertex points all
appear to move faster than light (see arrows).\endinsert

In this state the Cauchy surface is a single polygon, such that most of
its angles are very close to $180^\circ$, so it converges to a triangle
(sometimes an other simple shape). The sides move inwards with a
velocity exponentially approaching that of light. The particles (dots)
every now and then slip over the edges, after which they reappear at
one of the other image points of the vertex in question. This boosts
them so much that their velocity is much closer to that of light than
before, and the process is repeated an infinite number of times before
the universe has shrunk to a single point, at which it terminates its
existence.

It was found that a $g=0$ universe might begin with a Big Bang (the
time reverse of the above shrinking process) and either end expanding
forever or shrinking forever. This is sketched in Fig.~4a. If $g=1$ (a
torus), there are only two possibilities: either a Big Bang, or a Big
Crunch, but not both (Fig.~4b). We conjecture that at higher genus,
also an evolution from a shrinking mode into an expanding mode is
possible, but this was not checked explicitly.
\midinsert\epsffile{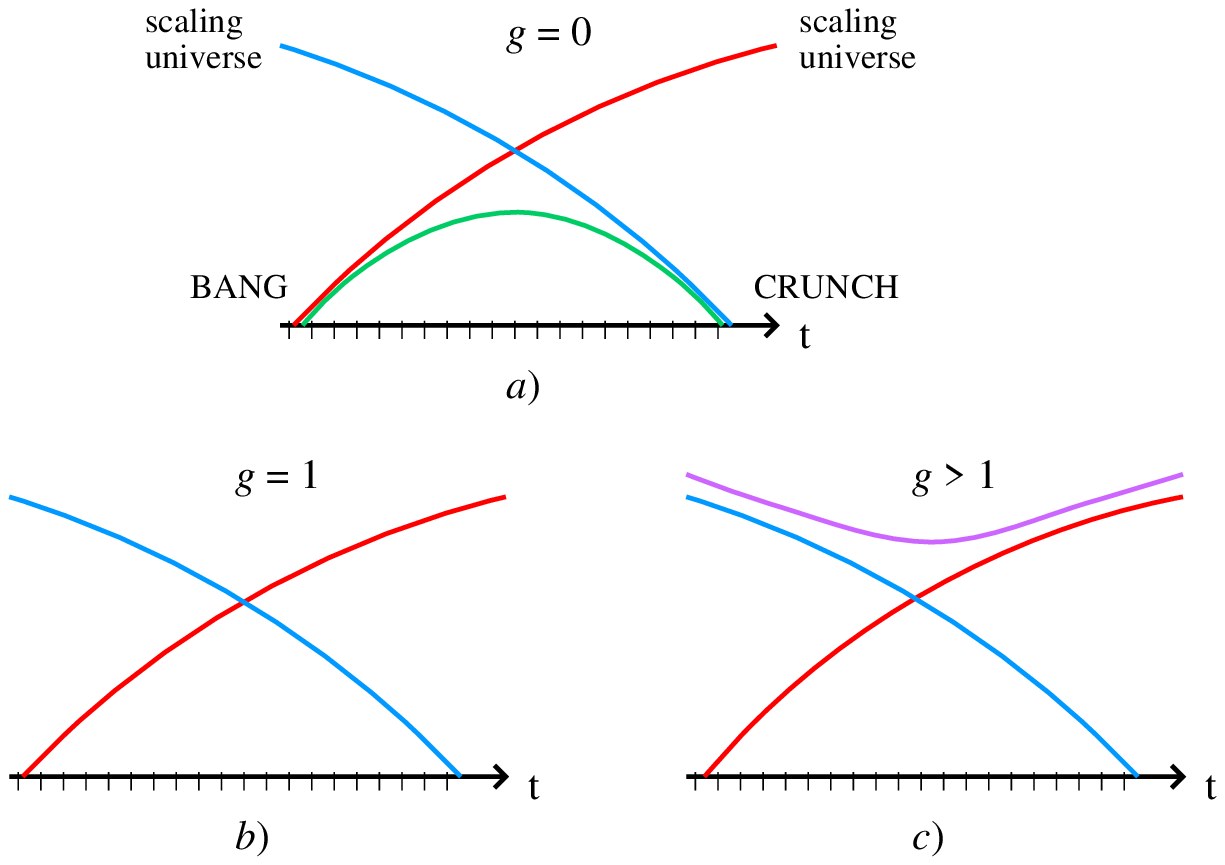}\narrower\ni{\bf Figure~4.} Evolving universes.
$a)$ an $S_2\times R_1$ topology; $b)$ if the topology is $S_1\times S_1
\times R_1$; $c)$ for higher topology (conjectured).\endinsert

As for the quantization of this model, there exist various opinions and
procedures. The Chern-Simons procedure as advocated by Carlip\ref4 and
Witten\ref5 does not indicate any discreteness in space and/or time.
Waelbroeck\ref6 claims that there are inequivalent quantization
procedures. In this author's opinion it is still not obvious whether
any of these procedures at all is completely consistent. Certainly one
would like to perform second quantization, so that in a limit where the
gravitational constant vanishes an ordinary scalar (or Dirac) field
theory emerges. This has never been demonstrated, and indeed, we find
that Hilbert spaces with transitions between states with different
particle numbers are difficult to construct. From Fig.~4, one suspects
that the evolution near a big Bang or a Big Crunc might violate
unitarity because there might not be acceptable states to evolve to or
from.

In the polygon representation, the most natural dynamical degrees of
freedom are the {\it lengths} $L_i$ of the edges of all polygons, and
their canonically conjugated variables, the Lorentz boost parameters
$\eta_i$ of Eqs. (1.2)--(1.4). If the Hamiltonian is taken to be
$$H_{\rm tot}\,=\,\sum_{{\rm particles}\,i}H_i\,+\,\sum_{{\rm
vertices}\,j}\w_j\,,\eqno(1.6)$$
with $H_i$ as described in (1.1)--(1.3) and $\w_j$ as in (1.5), then the
Poisson brackets
$$\{L_i,\, \eta_j\}\,=\,\d_{ij}\,,\eqno(1.7)$$
give the correct equations of motion:
$$\dot L_i\,=\,\{L_i,\,H\}\,.\eqno(1.8)$$

The fact that this gives time quantization\ref7 is then read off directly
from Eqs. (1.1)--(1.5), since the Hamiltonian consists exclusively of
angles.  The relevant operator one can construct directly is not $H$
but the time step operator $e^{\pm iH}$. In contrast, the lengths $L_i$
are {\it not} quantized, since their canonically conjugated variables
are hyperbolic angles, not real angles. If anything there is quantized,
it is the {\it imaginary parts\/} of $L_i$.

This situation changes radically if we use a different representation
of the particle system. It should be stressed that this is a change in
representation, not in the physical contents of the theory. We
introduce a {\it reference point}, the origin $\cal O$ of a coordinate
frame in 2-space, where the Lorentz frame will be kept fixed. Particles
can be reached from $\cal O$ via various different geodesics.  For each
particle $i$, at given time $t$, we take the shortest geodesic to that
particle, and use the coordinates $(x_i,\,y_i)$ of the particle seen
over this geodesic. Again, our 2-surface at given time is used as a
Cauchy surface, and we study its evolution. The same Hamiltonian is
used as before. Now we ask what the momentum variables are, conjugated
to $x_i$ and $y_i$. They form a vector $(p_{i,x},\,p_{i,y})$. The
length $p$ of this vector is found to be given by\ref8
$$p\,=\,\th\,\cos\m\,;\qquad \tan\th\,=\,\sinh\h\,.\eqno(1.9)$$
This is an angle! Consequently, the coordinates $x_i$ and $y_i$ are
quantized. Time remains quantized as it was before, since we did not 
change our Hamiltonian. Eq.~(1.3) turns into
$$\cos H\,=\,\cos\m\,\cos\th\,.\eqno(1.10)$$

We now refer to Ref\ref8 for a much more detailed exhibition of the
resulting lattice in 2+1 dimensional Minkowski space. A quick summary
is as follows. The angle $\th$, together with the orientation $\vv$ of
the momentum vector, form a compact 2-sphere. The space coordinates are
generated from the spherical harmonics on this 2-sphere, hence they are
represented by two integers $\ell$ and $m$. The {\it mass shell
condition}, Eq.~(1.10), is now a {\it difference equation\/} on
this lattice.  If $L_1,\,L_2,\,L_3$ are the usual angular momentum
operators on our spherical momentum space, the coordinates of one
particle can be identified as $$x\,=\,{L_2\over\cos\m}\,;\qquad
y\,=\,{-L_1\over\cos\m}\,;\qquad L\,=\,L_3\,.  \eqno(1.11)$$ Here, $L$
is the ordinary angular momentum in 2-space, and $\m$ is the particle
mass.  These could be seen as ``quantum coordinates":
$$\eqalign{[x,\,y]&\,=\,{i\over\cos^2\,\m}L\,,\cr
[L,\,x]&\,=\,iy\,,\cr [L,\,y]&\,=\,-ix\,.\cr}\eqno(1.12)$$

The difference equations for the wave function, as resulting from
Eq.~(1.10), is still second order in time. One can turn our wave
equation into a {\it Dirac equation\/} which is first order in time. The
Dirac particle has spin $\half$.
Second quantization should be performed by filling the Dirac sea, but a
difficulty encountered is that there will be two Fermi levels, of which
one carries negative energy particles.  We have no resolution of the
resulting problems at hand.

\secbreak
{\ni\bf 2. BLACK HOLE PHYSICS}\medskip
A direct generalization of the results of the previous chapter to 3+1
dimensions would lead to deceptive results. In 3+1 dimensions
space-time outside the matter sources is not flat; this would only be
if the matter sources could be taken to be stretches of rigid string
pieces. It would be highly preferable if we could derive certain features
concerning Planckian physics from facts out of everyday life, without
relying on any drastic assumptions. 

We now report that such a thing might well be possible. One well-known
fact in general relativity is that the gravitational force appears to
be unstable. given sufficient amounts of matter, gravitational
attraction can become so strong that colapse takes place, and no
classical variety of matter can withstand such a collapse. Indeed, if
the quantity of matter is large enough then during the collapse the
situation as seen by local observers may be quite normal and peaceful;
matter densities and temperatures could be those of ordinary water. 
According to the outside world however, a black hole is formed. As
long as one adheres to the formalisms of classical, that is, unquantized,
laws of physics, there is no contradiction anywhere. A black hole is
an interesting object, but we do not learn much from it about local
laws of physics.

Yet in a quantum theory what happens during gravitational collapse
turns out to be much more problematic and controversial. First of all
it is found that black holes will emit particles\ref9, and thereby
loose mass-energy.  Then one discovers that the laws of quantum field
theory at the local scale appear to be in conflict with the laws of
quantum mechanics for the black hole entire. Now we do not know whether
the black hole entire will obey ordinary laws of quantum mechanics, but
if it is allowed to decay into very tiny black holes that may pervade
the quantum vacuum state, we may arrive at a self-consistency problem.
Is or is not the small distance limit of our world quantum mechanical?
If not, how do we understand energy-momentum conservation and the
stability (and apparent uniqueness) of our vacuum?

The present author is investigating the train of thought following the
assumption that collapsing objects are still in complete agreement with
ordinary quantum mechanics (in particular there is no communication
with ``other universes" which would be tantamount to violation of
ordinary quantum determinism). The procedure has recently been laid
down precisely in our review paper\ref{10}, which we advise to be used
in conjunction with this paper. Here we will explain how ``quantization
of space and time" may follow from these considerations.
\def\Pl{{\rm Planck}}

Units are chosen such that 
$$\Gbar\,\df\,8\pi G\,=\,1\,,\eqno(2.1)$$
which gives us new Planck units of length, mass and energy: 
$$\eqalign{\Lbar_\Pl&\,=\,\sqrt{\hbar\,\Gbar\over c^3}\,=\,8.102\times
10^{-33}\rm cm\,,\cr
\Mbar_\Pl&\,=\,\sqrt{\hbar c\over\Gbar}\,=\,4.35 \m\rm g\,,\cr
\Ebar_\Pl\,=\,\Mbar_\Pl c^2&\,=\,\sqrt{\hbar c^5\over \Gbar}\,=\,2.39\times
10^{27} \rm eV\,.\cr}\eqno(2.2)$$

In its most elementary form, the {\it $S$-matrix Ansatz\/} for the
behavior of a black hole stipulates that, barring certain irrelevant
infra-red effects, the entire process of black hole creation and
subsequent evaporation can be viewed as a quantum mechanical scattering
event, to be described by a scattering matrix. In practice, for a given
black hole, it implies that the number of different possible states it
can be in is given by the exponent of the entropy $S=4\pi G M^2=\half
M^2$. This could be mimicked by a simple boundary condition near the
horizon (the ``brick wall"), forcing ingoing radiation to be bounced
back at a distance scale of the order of the Planck distance from the
horizon.

\midinsert\epsffile{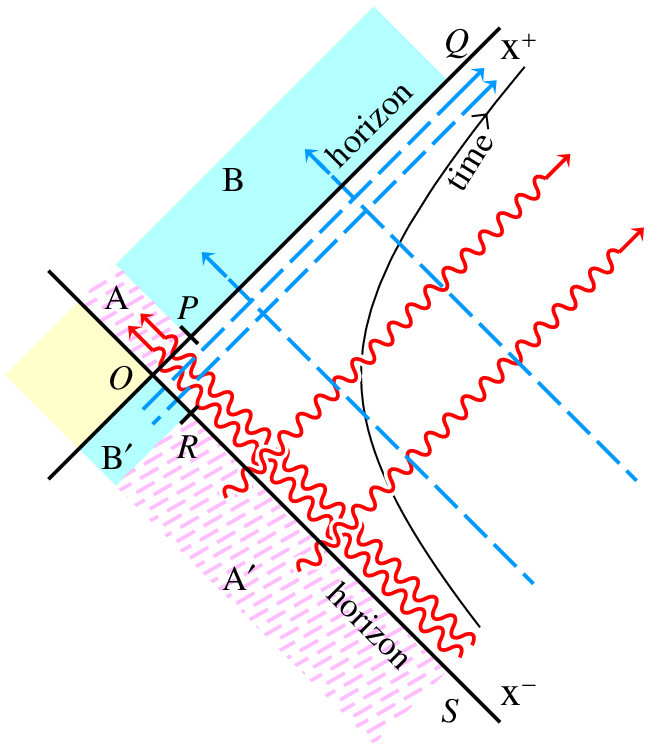} \narrower\ni{\bf Figure~5.}  Short
distance - large distance duality in the scattering matrix Ansatz.
Particles entering a black hole in $A$ will determine what comes out
from $A'$ (wavy lines); what enters at $B$ determines radiation from
$B'$ (dashed lines). The fields on the small region $OP$ are mapped as
fields on $RS$ and fields on $PQ$ are mapped onto $OR$.\endinsert

In terms of a local Rindler frame near the horizon, see Fig.~5, we
expect a mapping. All information passing the line $OQ$ in Fig.~5
should reemerge as information from the line $OS$. This implies that
the fields on $OQ$ determine the fields on $OS$. Such a mapping appears
not to exist in ordinary field theories in flat space-time. However,
one has to realize that the mapping relates distances shorter than the
Planck length (trans-Planckian distances) to distances large than the
Planck length (cis-Planckian distances). In Fig.~5, fields on the
trans-Planckian line $OP$ are mapped as fields on the cis-Planckian
line $RS$. Similarly, $PQ$ maps onto $OR$. This may be seen as a
long-distance-short distance duality not unlike $T$-duality as
discussed in string theories.

It is suspected that long the distance -- short distance duality
constraint should be imposed in all field theories in approximately
flat space-times, regardless whether the point $O$ (actually a
2-surface) acts as the intersection point of a futute horizon and a
past horizon, but we will concentrate on the case that there is a real
horizon.

\def\in{{\rm in}} 
In Ref\ref{10} it is explained how {\it interactions\/} between in- and
outgoing particles may restore a causal relationship that could
actually correspond to the mapping just described.  The most important
interaction here is the gravitational one. An ingoing particle with
momentum $p_\in$ causes a shift in the geodesics of outgoing particles.
This shift is usually in the inward direction, so it may be that
particles that were on their way out are moved back in again by am
ingoing particle. If the outging particles were represented as usual by
a Fock space, information loss would be unavoidable.

However, Fock space may have to be replaced by something else when it
comes to trans-Planckian (or near-Planckian) distance scales.  Two
particles that enter the horizon at the same anglular position $\tl
x=(\th,\,\vv)$ may have to be considered inseparable. Indicating the
coordinates of an outgoing particle as $(x^-,\,\tl x)$, we propose to
replace their Fock space by the set of observables $u^-(\tl x)$,
defined as 
$$u^-(\tl x)\quad\df\quad\Big\bra x_i^-(\tl x)\Big\ket_{\hbox{\smallrm
Average over}\atop \hbox{\smallrm all particles \smallit
i}}\,-\,x^-(\tl x)\bigg|_{\hbox{\smallrm Horizon}}\,.\eqno(2.3)$$
This is {\it one\/} observable at each transverse position $\tl x$.
Since there will always be particles at our side of the horizon, this
observable will continue to be observable regardless the amount of the
shift. Similarly, we have the observables $x^+(\tl x)$, referring to
the ingoing particles.

Being related to the actual position of the horizon, one might refer to
the operators $x^\m(\tl x)=\big(u^+(\tl x),\,u^-(\tl x),\,\tl x\big)$
as ``the shape of the horizon", more precisely, ``of the intersection
between past and future horizon." Later, we will replace the
independent coordinates $\tl x$ by a set of arbitrary coordinates
$\tl\s$, so that one has a sheet described as $x^\m(\tl\s)$.

According to the $S$-matrix Ansatz, $x^\m(\tl\s)$ contains {\it all\/}
information there is about the ingoing and outgoing states. Now, in the
conventional theory, this information is  contained by the fields in
the first quadrant. Thus we arrive at the important conclusion that
these fields can be {\it replaced\/} by the single (vector) function
$x^\m(\tl\s)$. This is what may be called {\it black hole
complementarity\/}\ref{11}, or, since we seem to have some sort of
projection of information in 3-space onto a two-dimensional
surface\ref{12}, the {\it holographic principle}.  It must be stressed,
however, that approximations were made; all non-gravitational forces
were neglected. Adding the electromagnetic force, for instance, yields
an additional component $x^5(\tl\s)$.  \secbreak

{\ni\bf QUANTIZATION OF SPACE AND TIME IN 3+1 DIMENSIONS}\medskip

The shift $\d x^-$ among the outgoing particles at transverse
coordinates $\tl x$ is proportional to the momentum $p_\in$ of the
ingoing particles at $\tl x'$:  $$\d x^-(\tl x)\,=\,\int\dd^2\tl
x'\,f(\tl x-\tl x')\,p_\in(\tl x')\,,\eqno(3.1)$$ where $f$ is a Green
function obeying 
$$\tl\pa^2f(\tl x)\,=\,-\,\d^2(\tl x)\,.\eqno(3.2)$$
If $x^+$ is the operator canonically conjugated to $p_\in=p_+$, one
would be tempted to write
$$\eqalign{[x^-(\tl x),\,x^+(\tl x')]\,=\,&f(\tl x-\tl x')\,[p_+(\tl x'),
\,x^+(\tl x)]\cr =\,&-\hbar if(\tl x-\tl x')\,.\cr}\eqno(3.3)$$
In case of many particles, labled by indices $i,\,j$:
$$[x^-_i(\tl x),\,x_j^+(\tl x')]\,\qu\,-\hbar if(\tl x-\tl x')\,\d_{ij}\,.
\eqno(3.4)$$
One then would have a ``quantum space-time", with beautifully
non-commuting coordinates. But this of course would be incorrect. Since
{\it all\/} ingoing particles interact gravitationally with {\it all\/}
outgoing ones, the Kronecker delta, $\d_{ij}$, should not be there. If
we had two ingoing particles, 1 and 2, that happen to be at the same
transverse position $\tl x$, then $x_1^+(\tl x)-x_2^+(\tl x)$ would be
an operator that commutes with everything, so that this ``observable"
would truly get lost in the black hole. We have to drop this
observable, as explained in the previous section, and we should work
exclusively with the horizon shape operator $x^\m(\tl\s)$ defined
there.

It is these operators that obey the commutation rule\ref{10, 13}
$$[x^-(\tl x),\,x^+(\tl x')]\,=\,-if(\tl x-\tl x')\,.\eqno(3.5)$$ Now
this equation has been derived for the case when one may neglect the
transverse components of the gravitational force. But if we
define\ref{10, 14} the surface orientation 2-form $W^{\m\n}=\dd
x^\m\wedge\dd x^\n$, or
$$W^{\m\n}(\tl\s)\,=\,-W^{\n\m}(\tl\s)\,=\,\e^{ab}\,\part{x^\m}{\s^a}
\part{x^\n}{\s^b}\,,\eqno(3.6)$$ 
we have, in the same approximation (where $\tl\s=\tl x$),
$$\sum_\m\big[W^{\m\a}(\tl\s),\,W^{\m\b}(\tl\s')\big]\,=\,\fractt12
\d^2(\tl\s-\tl\s')
\sum_{\m\n}\e^{\a\b\m\n}W^{\m\n}(\tl\s)\,.\eqno(3.7)$$
It is then argued that this equation, being Lorentz-invariant, should
continue to hold regardless the orientation of the gravitational shift.

\def\Kbar{\overline K}
Unfortunately, Eq.~(3.7) does not contain sufficient information to
find a representation of this algebra, since, at the left hand side,
there is still an index $\m$ that is summed over (without the summation one
gets non-local commutators). On the other hand, the $W^{\m\n}$ operators
overdetermine the surface $x^\m(\tl x)$. We therefore restrict ourselves
to its self-dual part $K_a(\tl\s)$, $a=1,2,3$. Defining $K$, and the anti-self-dual
part $\Kbar$, as
$$\eqalign{K_1(\tl\s)\,=\,iW^{23}+W^{10}\,,\qquad&\qquad\Kbar_1(\tl\s)\,=\,
-iW^{23}+W^{10}\,,\cr
K_2(\tl\s)\,=\,iW^{31}+W^{20}\,,\qquad&\qquad\Kbar_2(\tl\s)\,=\,
-iW^{31}+W^{20}\,,\cr
K_3(\tl\s)\,=\,iW^{12}+W^{30}\,;\qquad&\qquad\Kbar_3(\tl\s)\,=\,
-iW^{12}+W^{30}\,.\cr}\eqno(3.8)$$
we find the commutation rules
$$[K_a(\tl\s),\,K_b(\tl\s')]\,=\,i\e_{abc}K_c(\tl\s)\,\d^2(\tl\s-\tl\s')\,,
\eqno(3.9)$$ and similarly for the $\Kbar$. Mixed commutators of $K$
and $\Kbar$ are non-local however.

The operators $W$, $K$ and $\Kbar$ are distributions, so we want to
convolute them with test functions. It is convenient to take a test
function $\r(\tl\s)$ with the property $\r^2=\r$, which means that
$\r=1$ within some region in $\tl\s$ space and $\r=0$ in its
complement. Let $D$ be the domain where $\r=1$. Then
$$W^{\m\n}(D)\,\df\,\int\dd^2\tl\s\,g(\tl\s)W^{\m\n}(\tl\s)\,=\,
\int_D\dd^2\tl\s\, W^{\m\n}(\tl\s)\,=\,\oint_{\d D}x^\m\dd
x^\n\,.\eqno(3.10)$$ Defining $L_a(D)=\int_D K_a(\tl\s)\dd^2\tl\s$, we
find that these obey the commutation rules of angular momenta:
$$[L_a(D),\,L_b(D)]\,=\,i\e_{abc}L_c(D)\,.\eqno(3.11)$$ Thus, if we
divide the $\tl\s$-plane up in domains $D$, then we have a {\it
discrete representation\/} of our algebra, formed by the quantum
numbers $\{\ell_D,\,m_D\}$ for all domains. If two domains are combined
into one then $L(D_1+D_2)=L(D_1)+L(D_2)$, according to the familiar
rules of adding angular momenta.

It appears that the states $$|\{\ell_D,\,m_D\}\ket\,,\eqno(3.12)$$ with
$$\ell\,=\,0,\,\half,\,1,\,\fract32,\dots,\qquad
m=-\ell,\,-\ell+1,\dots, \ell\,.\eqno(3.13)$$ have the kind of
degeneracy one would expect for a black hole with entropy proportional
to its surface area. The $S$-matrix Ansatz would demand a degeneracy
not much worse than this. It must be stressed, however, that (3.12), (3,13)
is not the only representation of our algebra. The operators $L_a$ are
not hermitean. Instead, we have $$L_a^\dagger\,=\,\overline
L_a\,,\eqno(3.14)$$ and consequently one cannot derive the usual
properties (3.13) of the quantum numbers $\ell$ and $m$. We do have,
from the definition (3.6),
$$\e_{\m\n\a\b}W^{\m\n}W^{\a\b}\,=\,0\,.\eqno(3.15)$$ from which it
follows that $$K^2(\tl\s)\,=\,\Kbar^2(\tl\s)\,,\eqno(3.16)$$ but this
does not directly lead to constraints on $\ell_D$ and $m_D$; for
instance, $\ell_D$ could easily be negative.  One does have the
orthonormality property $$\bra\{\overline\ell_D,\,\overline
m_D\}|\{\ell'_D,\,m'_D\}\ket\,=\,
\prod_D\d_{\overline\ell_D,\ell'_D}\d\low{\overline
m_D,m'_D}\,,\eqno(3.17)$$ where $\overline\ell_D$ and $\overline m_D$
refer to the representations of $\overline L_a(D)$.

Since the $S$-matrix Ansatz requires a finite degeneracy, we can now
ask ourseves what the consequences would be for the $K$ and $\Kbar$
operators if we do restrict ourselves to the representations (3.13). If
the operators $K_a$ differ only infinitesimally from $\Kbar_a$,
Eq.~(3.13) should still hold.  Thus, we want the real parts in
Eq.~(3.8) to be much bigger than the imaginary parts. If
$$x^0,\,x^3\,\gg \,x^1,x^2\,,\eqno(3.18)$$ then $\ell$ is real and
$m\approx\ell\gg 0$. Note that in this case we have a {\it timelike\/}
surface, whereas the horizon surface that we started off with was
spacelike. We suspect that what (3.13) is really telling us is that the
smallest domains must be timelike surface elements, and that the
spacelike horizon can be considered to be a globally spacelike
patchwork of many such timelike pieces.

However, the constraint (3.13) is not yet fully guaranteed by (3.18).
It is better to postulate for each domain $D$
$$|\d x^0|\,\gg \,|\d x_i|\,,\qquad i=1,2,3\,.\eqno(3.19)$$
Again, this describes timelike ``string worldsheets" joined together
to form the horizon. A more precise interpretation is as follows.

We may {\it choose\/} the shapes of the domains $D$. For instance, we
may choose the time intervals $\d x^0$, and draw the domains as rectangles
(Fig.~6a). If we choose these to be an
integer multiple of a quantum $\D t$, then time is quantized. The time
quantum $\D t$ is arbitrary, but as for now we choose it to be much
bigger than the Planck time.

In this case, for small enough domains, 
$$L_a(D)\,\approx\,\overline L_a(D)\,\approx\,\oint_{\d D} x^a\dd x^0
\,=\,\D t\cdot\big(x_a(2)-x_a(1)\big)\,,\eqno(3.20)$$
where $x_a(1)$ is the average value of $x_a$ at one edge of the rectangle
and $x_a(2)$ the average value at the other side.
Writing $\d x_a=x_a(2)-x_a(1)$, we find
$$L_a(D)\,=\,\D t\cdot\d x_a\,.\eqno(3.21)$$
Consequently, $\d x_a$ are quantized in multiples of
$$\D x\,=\,\fractt12/\D t\,.\eqno(3.22)$$
Putting the units back in, we have
$$\D t\cdot\D x\,=\,4\pi G\,.\eqno(3.23)$$
The resulting ``string" is pictured in Fig.~6b. We note that the string bits
are vectors in 3-space obeying the quantization rules of angular momenta.

\midinsert\epsffile{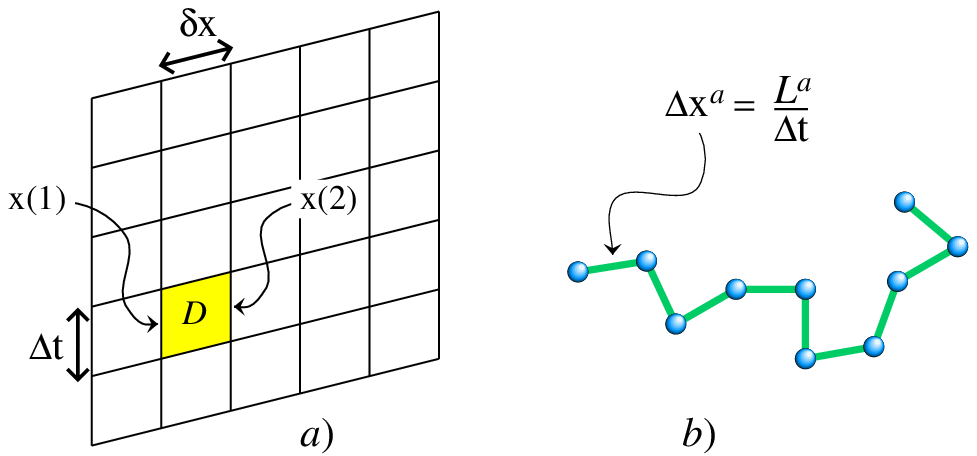}\narrower\ni{\bf Figure 6.} $a$) Timelike
segment of the horizon, divided into domains $D$. \ $b$) At given time
$t$ the string consists of pieces quantized in units of $\D x$,
obeying the commutation rules of angular momenta.\endinsert

Apparently, space-time now forms a lattice. Note that we did {\it
not\/} derive equations of motion for this string, whose target space
appears to be a quantum space-time, much like in the 2+1 dimensional
case. Note also that the string bit vector elements at a given time {\it
commute} with the string bit vector elements at other times, unlike the
situation in ordinary field theories. Our space-time quantization rules
have much in common with the surface area quantization rules suggested by
Bekenstein and Mukhanov\ref{15}, for example, but are more detailed. 

An interesting consequence of Eq.~(3.23) is that the Hamiltonian will
be limited to the region $0\le H<2\pi/\D t=\D x/2G$. Apparently,
gravitational disturbances of 3-space then always remain within one
space quantum away from flat space.  It goes without saying that the
question exactly how all this has to be combined in a more
comprehensive theory remains to be studied.

\secbreak{\bf\ni REFERENCES}
\item{1.} H. Waelbroeck, {\it Class. Quantum Grav.} {\bf 7} (1990) 751;
{\it Phys. Rev. Lett.} {\bf 64} (1990) 2222; {\it Nucl. Phys.} {\bf B
364} (1991) 475.
\item{2.} G.~'t~Hooft, {\it Class. Quantum Grav.} {\bf 9} (1992) 1335;
{\bf 10} (1993) S 79.
\item{3.} G. 't Hooft,  {\it Class. Quantum Grav.} {\bf 10} (1993) 1023.
\item{4.}  S. Carlip, {\it Nucl. Phys.} {\bf B324} (1989) 106; and in: ``{\it
Physics,  Geometry and Topology\/}", NATO ASI series B, Physics, Vol.
{\bf 238}, H.C.  Lee ed., Plenum 1990, p. 541;  S. Carlip, {\it Six
ways to quantize (2+1)-dimensional gravity}, Davis Preprint UCD-93-15,
gr-qc/9305020. 
\item{5.}  E. Witten, {\it Nucl. Phys.} {\bf B311} (1988) 46; {\it see also\/}
 G. Grignani, {\it 2+1-dimensional gravity as a gauge
theory of the Poincar\'e group}, Scuola Normale Superiore, Perugia,
Thesis 1992-1993. 
\item{6.}  H.~Waelbroeck and F.~Zertuche, {\it Phys. Rev.} {\bf
D50} (1994) 4966; H.~Waelbroeck, pers. comm.
\item{7.}  G. 't Hooft, {\it Class. Quantum Grav.} {\bf 10} (1993) 1653
(gr-qc/9305008). \br 
H.~Waelbroeck and J.A.~Zapata, {\it 2+1 Covariant Lattice
Theory and 't~Hooft's Formulation}, Pennsylvania State Univ. prepr.
CGPG-95/8 (gr-qc/9601011). \br See also: A.P. Balachandran and L. Chandar, {\it
Nucl. Phys.} {\bf B 428} (1994) 435.
\item{8.} G.~'t~Hooft, {\it Class. Quantum Grav.} {\bf 13} (1996) 1023
(gr-qc/9601014).
\item{9.}  S.W. Hawking, Commun. Math. Phys. {\bf 43} (1975) 199; J.B.
Hartle and S.W. Hawking, Phys. Rev. {\bf D13} (1976) 2188.
\item{10.} G.~'t~Hooft, ``The Scattering matrix Approach for the Quantum
Black Hole'' (gr-qc/9607022).
\item{11.} L.~Susskind, L.~Thorlacius and J.~Uglum, Phys. Rev. D {\bf 48} 
(1993) 3743 (hep-th/9306069). 
\item{12.} G.~'t~Hooft,  ``Dimensional Reduction in Quantum Gravity",
in {\it Salamfestschrift: a collection of talks}, World Scientific Series in
20th Century Physics, vol.~4, ed. A.~Ali, J.~Ellis and S.~Randjbar-Daemi 
(World Scientific, 1993), THU-93/26,  gr-qc/9310026;\br
L.~Susskind, ``The world as a hologram", J. Math. Phys. {\bf 36} (1995) 6377, hep-th/9409089;\br
S. Corley and T. Jacobson, ``Focusing and the holographic hypothesis", gr-qc/9602031.
\item{13.}  G. 't Hooft,  Physica Scripta, {\bf  T15} (1987) 143-150;
{\it id.}  Nucl. Phys.  {\bf B335} (1990) 138; Physica
Scripta T {\bf 36} (1991) 247. 
\item{14.} G.~'t~Hooft, ``The Black Hole Horizon as a Quantum Surface"  
{\it Physica Scripta} {\bf T36} (1991) 247-252.  
Also published in: {\it The Birth and Early Evolution of Our Universe}, 
Proceedings of Nobel Symposium 79, Gr\"aftavallen, Sweden, June 11-16, 1990.
(Eds.: J.S. Nilsson, B. Gustafsson and B.-S. Skagerstam.) World Scientific,
Singapore (1991).
\item{15.} J.D.~Bekenstein and V.F.~Mukhanov, Phys. Lett. B {\bf 360} (1995) 7.

\bye